\documentclass[aps,prb,twocolumn,superscriptaddress,10pt,reprint,floatfix]{revtex4-2}

\usepackage{amsmath,amssymb}
\usepackage{graphicx}

\usepackage[usenames,dvipsnames]{xcolor}
\usepackage[percent]{overpic}
\usepackage[pdftex,
    pdflang={en-US},
    colorlinks=true,
    urlcolor=blue,
    citecolor=red,
    pdfauthor={Pontus Laurell, Bradraj Pandey, Gabor Halasz, Satoshi Okamoto, Gonzalo Alvarez, and Elbio Dagotto},
    pdftitle={Pairing tendencies in the doped Kitaev-Heisenberg model}
    ]{hyperref}
\usepackage[stretch=10,shrink=10]{microtype}

\makeatletter
\def\maketitle{
	\@author@finish
	\title@column\titleblock@produce
	\suppressfloats[t]}
\makeatother

\begin{document}
	
\title{Pairing tendencies in the doped Kitaev-Heisenberg model}

\author{Pontus Laurell}
\thanks{These authors contributed equally to this work}
\altaffiliation{Present address: Department of Physics and Astronomy, University of Missouri, Columbia, Missouri 65211, USA}
\email{plaurell@missouri.edu}
\affiliation{Department of Physics and Astronomy, University of Tennessee, Knoxville, Tennessee 37996, USA}

\author{Bradraj Pandey}
\thanks{These authors contributed equally to this work.} \email{bradraj.pandey@gmail.com}
\affiliation{Department of Physics and Astronomy, University of Tennessee, Knoxville, Tennessee 37996, USA}
\affiliation{Materials Science and Technology Division, Oak Ridge National Laboratory, Oak Ridge, Tennessee 37831, USA}

\author{G\'abor B. Hal\'asz}
\affiliation{Materials Science and Technology Division, Oak Ridge National Laboratory, Oak Ridge, Tennessee 37831, USA}
\affiliation{Quantum Science Center, Oak Ridge, Tennessee 37831, USA}

\author{Satoshi Okamoto}
\affiliation{Materials Science and Technology Division, Oak Ridge National Laboratory, Oak Ridge, Tennessee 37831, USA}

\author{Gonzalo Alvarez}
\affiliation{Computational Sciences and Engineering Division, Oak Ridge National Laboratory, Oak Ridge, Tennessee 37831, USA}

\author{Elbio Dagotto}
\affiliation{Department of Physics and Astronomy, University of Tennessee, Knoxville, Tennessee 37996, USA}
\affiliation{Materials Science and Technology Division, Oak Ridge National Laboratory, Oak Ridge, Tennessee 37831, USA}

\begin{abstract}
We study the impact of hole-doping on the Kitaev-Heisenberg model on the honeycomb lattice. We investigate the pairing tendencies and correlation functions in the framework of a $t-J-K$ model using density matrix renormalization group calculations on three-leg cylinders. In the case of the pure Kitaev model, which realizes a quantum spin-liquid phase at half-filling, we find that binding of two holes only occurs at low values of the hopping, where the holes are slow. We have theoretically verified that pair formation occurs in the limit of immobile holes, where the pure Kitaev model remains exactly solvable. 
When we instead fix the hopping at an intermediate, more realistic, value, and vary the Heisenberg and Kitaev interaction strengths, we find pairing tendencies only in the N\'eel phase. This is in contrast to prior mean-field calculations, highlighting the importance of accounting for the kinetic energy of dopants in generalized Kitaev models. Interestingly, we also find signatures of pair-density wave formation over the studied range of model parameters, namely a periodic modulation of the charge density as well as the spin-spin and pair-pair correlations in real space. Moreover, we present a comparative study of the different correlations as a function of doping. We finally discuss the potential for experimentally observing the studied physics in quantum materials and heterostructures.
\end{abstract}


\maketitle

\section{Introduction}

There is a long-standing interest in doping quantum spin liquid (QSL) states \cite{Anderson_1987, PhysRevB.35.8865, Laughlin_1988, RevModPhys.63.1, PhysRevLett.70.493, PhysRevB.49.12058, RevModPhys.66.763, doi:10.1126/science.271.5249.618, doi:10.1126/science.289.5478.419, RevModPhys.78.17, Lee2007, Peng2021a, Arovas2022, PhysRevLett.130.136003}, \emph{i.e.}, phases of matter which occur in interacting quantum spin systems and are characterized by topological properties, exotic excitations, and the absence of magnetic ordering even at zero temperature \cite{Savary2017, Knolle2019, Broholm2020}. This direction dates back to the proposal \cite{Anderson_1987} that resonating valence bond (RVB) states are relevant to the description of unconventional superconductivity in, \emph{e.g.}, the cuprates. The RVB wave function involves a superposition of singlet states, such that the introduction of dopants into this liquid background leads to spin-singlet Cooper pair formation \cite{Anderson2004}. It is natural to consider other liquid backgrounds, as they can potentially lead to more exotic forms of pairing.

A particularly interesting QSL is realized by Kitaev's honeycomb spin-$1/2$ model \cite{Kitaev2006}. 
This exactly solvable model features competing bond-dependent interactions and a QSL ground state with two types of fractionalized excitations: gapless Majorana fermions and gapped visons (\emph{i.e.}, emergent gauge fluxes) \cite{Hermanns2018}. If time-reversal symmetry is broken by, \emph{e.g.}, a magnetic field, the Majorana fermions are gapped out, and the visons acquire non-Abelian particle statistics. Hence, Kitaev proposed \cite{Kitaev2006} his Hamiltonian as a model for topological quantum computing \cite{Kitaev2003, RevModPhys.80.1083}. Following the realization that the bond-dependent interactions of the Kitaev model can emerge in transition metal systems with edge-sharing octahedra and strong spin-orbit coupling \cite{PhysRevLett.102.017205}, such as Na$_2$IrO$_3$ \cite{PhysRevB.82.064412, PhysRevB.83.220403, PhysRevLett.108.127203, PhysRevLett.108.127204, PhysRevLett.110.097204, Chun2015} and RuCl$_3$ \cite{PhysRevB.90.041112, Banerjee2016, Banerjee2017, Do2017, 10.1038/s41467-017-01177-0, Laurell2020, PhysRevResearch.2.033011, PhysRevResearch.4.L022061}, there has also been a significant effort to find and characterize candidate Kitaev materials \cite{Takagi2019, doi:10.7566/JPSJ.89.012002, Trebst2022, Rousochatzakis_2024}.

However, since materials tend to have additional hopping paths and non-ideal bond geometries, their spin Hamiltonians include additional interaction terms beyond the pure Kitaev model description, inducing magnetic order at low temperature as is observed experimentally. The exact solution is no longer valid in the presence of these interactions, making approximate theories and numerical many-body techniques indispensable. The simplest extension of Kitaev's Hamiltonian is the Kitaev-Heisenberg model \cite{PhysRevLett.105.027204}, which also incorporates a spin-isotropic Heisenberg exchange term. The phase diagram (see Figure~\ref{fig:model}) and properties of the Kitaev-Heisenberg model at half-filling are well-established at this point
\cite{PhysRevLett.105.027204, PhysRevB.83.245104, PhysRevB.84.100406, PhysRevB.86.224417, PhysRevB.87.064508, PhysRevLett.110.097204, PhysRevLett.112.077204, PhysRevB.88.024410, PhysRevB.91.054401, PhysRevB.92.020405, PhysRevLett.119.157203, PhysRevB.97.134432}, but the behavior upon doping remains uncertain.

We focus here on hole doping, the impact of which can depend on both the concentration and kinetic energy of the dopants. The propagation of a single fast hole (with $t>K$, where $t$ is the hopping and $K$ is the Kitaev interaction strength) in a $t-J$-like generalization of the Kitaev-Heisenberg model was studied using exact diagonalization in Refs.~\cite{PhysRevLett.111.037205, PhysRevB.90.024404}. Signatures of coherent quasiparticle propagation were found in spectral functions for the N\'eel phase, similar to the behavior in $t-J$ models for cuprates \cite{RevModPhys.66.763} and consistent with prior work on the honeycomb $t-J$ model \cite{PhysRevB.73.155118}. However, the sublattice structure of the zigzag and stripy phases was found \cite{PhysRevLett.111.037205, PhysRevB.90.024404} to hinder coherent propagation, leading to ``hidden quasiparticles'', whereas holes in the Kitaev spin liquid phase propagated entirely incoherently. A self-consistent Born approximation study \cite{Wang_2018_Kitaevhole} obtained qualitatively similar results. In stark contrast, an analytical variational calculation for slow holes ($t\ll K$) at the Kitaev point (where slow holes only couple to the gapless Majorana modes) found that a single slow hole propagates coherently \cite{PhysRevB.94.235105}. These contrasting findings strongly suggest that the behavior of dopants depends nontrivially on both the magnetic background and on their kinetic energy. Two recent density matrix renormalization group (DMRG) papers \cite{Kadow_2024, Jin2024} about the dynamic response to injection of a single hole into four-leg cylinders reinforce this distinction between fast and slow holes in the case of antiferromagnetic Kitaev interactions \cite{Kadow_2024}, and also find a strong susceptibility to Nagaoka ferromagnetism for ferromagnetic Kitaev interactions \cite{Kadow_2024, Jin2024}.

The effects of higher doping levels $\delta$ have been studied using a variety of methods. 
A fermionic ``dopon'' mean-field theory was constructed at the Kitaev point \cite{PhysRevLett.108.227207}. This theory predicts a Fermi liquid phase with well-defined quasiparticles for $t\delta\ll K$, where the kinetic energy is small enough to leave the Kitaev spin liquid background unchanged, and also predicts $p$-wave superconductivity at large doping levels. On the other hand, slave-boson mean-field theory studies
\cite{PhysRevB.85.140510, PhysRevB.86.085145, PhysRevB.87.064508, PhysRevB.97.014504}
have reported multiple superconducting phases in the doped Kitaev-Heisenberg model. 
For FM Kitaev interaction ($K\leq 0$) and AFM Heisenberg interaction ($J
\geq 0$), Refs.~\cite{PhysRevB.85.140510, PhysRevB.86.085145} found two distinct $p$-wave superconducting phases emerging from the Kitaev point upon doping. $d$- and $s$-wave phases were also reported at higher $J$. 
Ref. \cite{PhysRevB.87.064508} clarified the nature of the $p$-wave phases and extended the calculation to all signs of $J$ and $K$, finding distinct differences between FM and AFM Kitaev couplings. For $K>0$, a larger area of the phase diagram results in $d+id$ pairing, whereas in both cases sufficiently strong FM Heisenberg interaction leads to a ferromagnetic phase. Ref.~\cite{PhysRevB.97.014504} analyzed an extended model that also includes off-diagonal $\Gamma$ interactions. 

A functional renormalization group (fRG) study \cite{PhysRevB.90.045135} found a remarkably similar $K<0$, $J\geq 0$ phase diagram to Refs.~\cite{PhysRevB.85.140510, PhysRevB.87.064508} when accounting only for particle-particle contributions to the (truncated) fRG flow equations. (We note that, while Refs.~\cite{PhysRevB.85.140510, PhysRevB.90.045135} both used a parametrization in which $|K|/t=1$, Ref.~\cite{PhysRevB.87.064508} had $|J|+|K|=2t$, complicating direct comparisons.) 
When particle-hole contributions are also included in the fRG calculations, the phase diagrams for $K<0$, $J\geq 0$ and $K>0$, $J\leq 0$ were found to include sizeable regions with antiferromagnetic and charge-density wave instabilities, respectively \cite{PhysRevB.90.045135}. This highlights the importance of beyond-mean-field effects in the physics of doped Kitaev-Heisenberg models. It is worth noting that the fRG calculations themselves include other approximations, such as truncation of the flow equations, the impact of which is not easily quantifiable.

Recently, the antiferromagnetic Kitaev point ($J=0$, $K/t=1/3$) was studied using large scale DMRG calculations on cylinders of three and four legs \cite{Peng2021}. Intriguingly, Ref. \cite{Peng2021} found that the leading pairing instability in the lightly doped case is of the sought-after pair-density type \cite{RevModPhys.87.457, doi:10.1146/annurev-conmatphys-031119-050711}. However, the pair-pair correlations decay rapidly, making the true long-distance behavior uncertain. Ref. \cite{Jin2024} found that the decay is exponential at the ferromagnetic Kitaev point for a large value of the hopping ($J=0$, $K/t=-1/20$). These results show the potential for DMRG calculations to elucidate the effects of doping, at least for finite-size Kitaev systems.

We thus undertake here a DMRG study of the hole-doped Kitaev-Heisenberg model in three regions of parameter space: (i) $K\geq 0$, $J\leq 0$, (ii) $K\geq 0$, $J\geq 0$, and (iii) $K\leq 0$, $J\geq 0$. We report a strong $t$-dependence of the binding energy at the Kitaev point, which supports the distinction between fast and slow holes previously made on analytical and computational grounds  \cite{PhysRevB.94.235105, Kadow_2024}. For fixed hopping $t=1$, we find evidence of negative binding energy, a prerequisite for pair formation, only in the N\'eel phase. We find signals of pair-density-wave precursors throughout the $K\geq 0$, $J\geq 0$ region, suggesting such correlations are more prevalent than previously expected. We find that the slow-hole regime is most promising for pairing inside, or proximate to, the spin liquid phase and discuss its relevance to heterostructures. We also calculate the binding energy at the Kitaev point in the spin-vacancy limit, $t=0$, using the exact solution.

The organization of the manuscript is as follows: Section \ref{sec:model} introduces the doped Kitaev-Heisenberg model and its phase diagram at half-filling, and also provides details about our numerical methods. Section \ref{sec:results} describes our results for the binding energy (Sec. \ref{sec:results:binding}), pair-pair correlations (Sec. \ref{sec:results:pairpair}), tendencies towards formation of exotic pair-density waves in the N\'eel and spin-liquid phases with doping (Sec. \ref{sec:results:PDW}), and the contrasted spin-spin, charge-charge, and pair-pair correlation functions in the N\'eel phase (Sec. \ref{sec:results:comparison}). Section \ref{sec:discussion} provides an intuitive picture for hole-pair formation in the model and presents explicit binding energy results for the pure Kitaev model in the limit of static holes. Also discussed are consequences of our findings for realistic materials. Finally, Section \ref{sec:conclusion} presents our conclusions.

\section{Spin model and methods}\label{sec:model}
\begin{figure}[ht]
    \includegraphics[width=0.8\columnwidth]{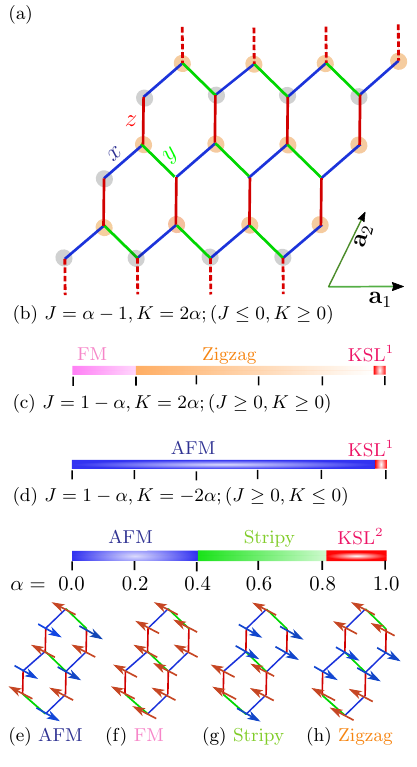}
    \caption{The Kitaev-Heisenberg model.
    (a) Schematic representation of a three-leg cylinder on the honeycomb lattice. The distinct $x$, $y$, and $z$ bonds are denoted by solid blue, green and red lines, respectively. Dashed red lines indicate $z$ bonds between sites connected by the periodic boundary conditions of the cylinder.
    (b) The phase diagram for $J\leq 0$ and $K\geq 0$: we find ferromagnetic (FM), Zigzag, and Kitaev-spin liquid (KSL$^1$) phases varying $\alpha$.
    (c) For  $J\geq 0$ and $K\geq 0$ we find antiferromagnetic N\'eel (AFM)
  and Kitaev spin liquid phases. 
    (d) The phase diagram for $K\leq 0$ and  $J\geq 0$: we find AFM, stripy and Kitaev spin liquid (KSL$^2$) phases. 
    Pictorial representations are provided for the (e) AFM, (f) FM, (g) stripy, and (h) zigzag  phases.}
\label{fig:model}
\end{figure}

The Hamiltonian for a doped Kitaev-Heisenberg model on the honeycomb lattice can be written as
\begin{align}
    H &= K \sum_{<i,j>}S_i^{\gamma}S_j^{\gamma}+J\sum_{\langle i,j\rangle} \mathbf{S}_i\cdot \mathbf{S}_j \nonumber\\
    &-t\sum_{<i,j>,\sigma}\left( c^{\dagger}_{i\sigma}c_{j,\sigma}+\mathrm{H.c.}\right),   \label{eq:ham}
\end{align}
where $K$ is the Kitaev interaction strength and $J$ is the Heisenberg exchange strength. $\mathbf{S}_i = \left( S_i^x, S_i^y, S_i^z\right)$ is the vector of spin-$1/2$ operators on site $i$, and sums are taken over nearest neighbors (NN). In the first term, $\gamma\in \left\{ x,y,z\right\}$ labels the three distinct types of NN bonds shown in Fig. \ref{fig:model}(a). The last term describes the effect of doping, with $t$ being the hopping parameter for the doped carriers, $ c^{\dagger}_{i\sigma}$ being the electron creation operator with spin $\sigma$, and $\mathrm{H.c.}$ denoting Hermitian conjugate. 
The total number of sites in a finite-size cluster is $N=2\times L_x \times L_y$, where $L_x$ and $L_y$ represent the number of repetitions of the lattice vectors, $\mathbf{a}_1$ and $\mathbf{a}_2$, respectively.

We note that YbCl$_3$ \cite{Sala2021} provides an excellent realization of the NN Heisenberg model (i.e., Eq.~\eqref{eq:ham} at half-filling with $K=0$). The extent to which the Kitaev-Heisenberg model is realized in materials remains an open question because, generically, additional off-diagonal and further-range interactions can be important \cite{PhysRevLett.112.077204, PhysRevB.93.214431, PhysRevB.100.075110, Rousochatzakis_2024}. However, it represents a minimal model for theoretical study, that captures both the role of the Kitaev coupling and integrability-breaking, which is why we adopt it here.

In order to numerically solve the above Hamiltonian, we use the DMRG method \cite{PhysRevLett.69.2863, PhysRevB.48.10345} as implemented in the DMRG++ software \cite{Alvarez2009}. 
The lattice cylinder geometry is described in Fig. \ref{fig:model}(a). 
We use 
open boundary conditions along the long direction ($\mathbf{a}_1$) and periodic boundary conditions along the short direction ($\mathbf{a}_2$). 
We consider mainly three-leg ladders with system size up to $2\times L_y\times L_x=2\times 3 \times 12$
(we have also studied a few cases using a four-leg ladder). 

Following Ref. \cite{PhysRevLett.105.027204}, we parametrize the Kitaev $K$ and Heisenberg $J$ couplings with a single parameter $ 0 \le \alpha \le 1$ 
for three different cases as: 
\begin{enumerate}
    \item $J=\alpha-1,\, K=2\alpha;\quad (J\leq 0,\, K\geq 0)$
    \item $J=1-\alpha,\, K=2\alpha;\quad (J\geq 0,\, K\geq 0)$
    \item $J=1-\alpha,\, K=-2\alpha;\quad (J\geq 0,\, K\leq 0)$
\end{enumerate}
The ground state phase diagrams at half-filling (i.e. one electron per site) obtained from our DMRG calculations are shown in Figs. \ref{fig:model}(b)-(d). For $J\leq 0$, $K\geq 0$ [Fig. \ref{fig:model}(b)] one finds the FM state [Fig. \ref{fig:model}(f)] at low $\alpha$. The competition between $J<0$ and $K>0$ results in a large regime with zigzag ordering [Fig. \ref{fig:model}(h)], whereas the AFM Kitaev spin liquid (KSL$^1$) phase without any magnetic order is obtained at large $\alpha$. 

For $J\geq 0$, $K\geq 0$ [Fig. \ref{fig:model}(c)] the AFM N\'eel [Fig. \ref{fig:model}(e)] is found over a large range of $\alpha$ values, whereas the AFM Kitaev spin liquid phase is found only for a small range of $\alpha$. For $J\geq 0$, $K\leq 0$ [Fig. \ref{fig:model}(d)], the N\'eel phase survives for $\alpha \leq 0.4$, while $0.4 \le \alpha \le 0.8$ produces the stripy phase [Fig. \ref{fig:model}(g)]. Large $\alpha$ results in the FM Kitaev spin liquid (KSL$^2$) phase. We note that the results for this region can be directly compared with the phase diagram obtained in Ref. \cite{PhysRevLett.105.027204}.

To study superconducting and magnetic properties of the doped Kitaev-Heisenberg model, we remove $n_h$ electrons from (or add $n_h$ holes to) the half-filled system. We constrain the Hilbert space of a single site to contain either zero or one electrons, i.e., double occupancy is disallowed, making our doped system a $t-J$-like model with regards to its Hilbert space. We calculate the pair binding energy and various types of correlation functions (pair-pair, spin-spin, and charge-charge), to be defined below. For the majority of the calculations we kept up to $m=5600$ DMRG states, finding truncation errors $\epsilon \sim 10^{-5}$. Some binding energy calculations used up to $m=7200$ states. See the supplemental material for additional details \cite{SuppMat}.

\section{Results}\label{sec:results}
We present results for the binding energy of two holes as well as correlation functions, and investigate the possibility of pair density wave states.

\subsection{Pairing tendencies}\label{sec:results:binding}
\begin{figure}[ht]
\centering
\includegraphics*[width=\linewidth]{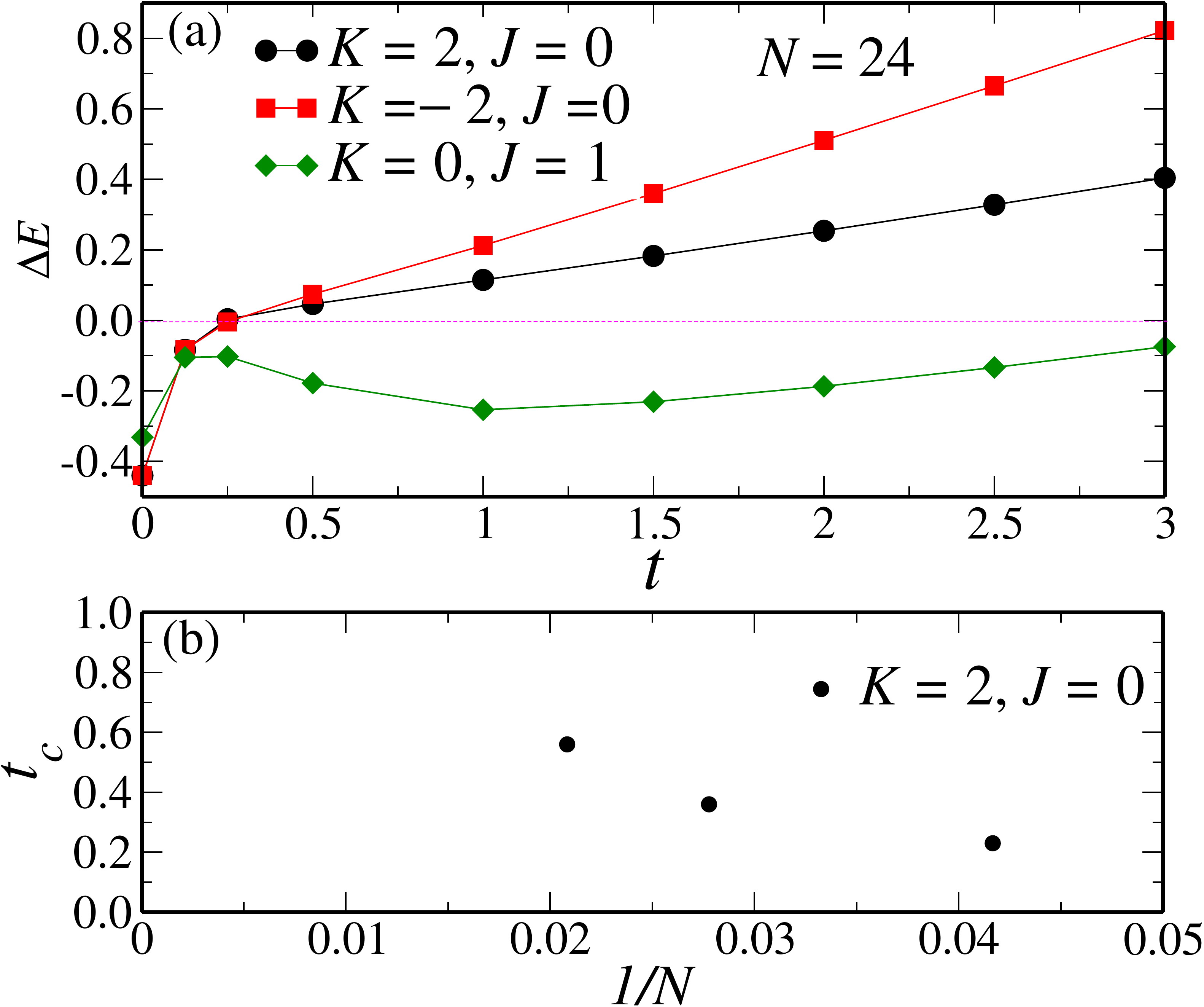}
\caption{(a) Impact of the kinetic energy of dopants on pair formation. The binding energy $\Delta E$ is plotted as a function of the hole hopping amplitude $t$ for (i) the AFM Kitaev point with $K=2,\,J=0$, (ii) the FM Kitaev point with $K=-2,\,J=0$, and (iii) the AFM Heisenberg point with $K=0,\, J=1$  for a fixed system size with $N=24$ sites. (b) The crossover hopping $t_c$, defined as the value of $t$ where $\Delta E$ changes sign, is plotted as a function of the inverse system size $1/N$ at the AFM Kitaev point.
}
\label{fig:binding:hopping}
\end{figure}

    \begin{figure}[!ht]
    \begin{overpic}[width=0.92\columnwidth]{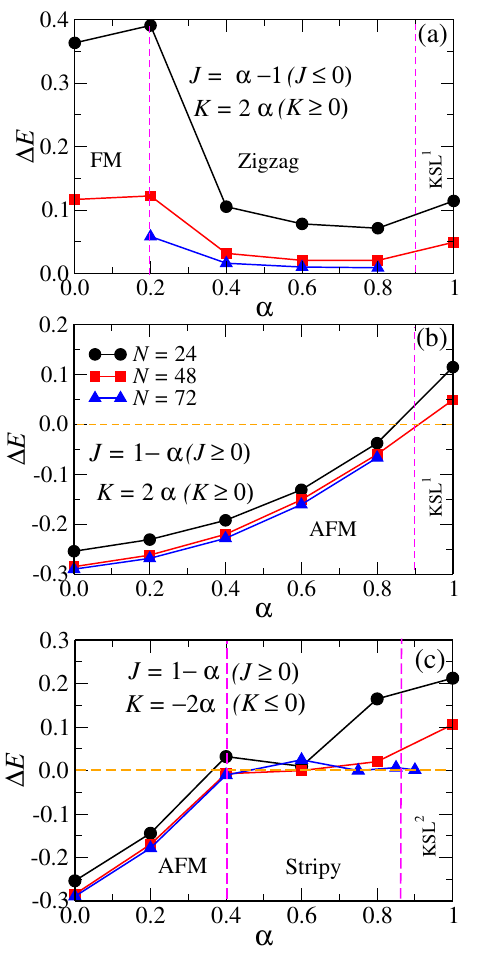}
    \end{overpic}
    \caption{Binding energy $\Delta E$ vs $\alpha$ for cluster sizes $N=2\times 3 \times 4$,  $N=2\times 3 \times 8$ and  $N=2\times 3 \times 12$.
 (a) For $J = \alpha-1$ and $K =2\alpha$, the binding energy is positive for all three phases, including the ferromagnetic (FM), zigzag and 
Kitaev spin liquid (KSL) phases. (b) For $J = 1- \alpha$ and $K =2\alpha$, the binding energy is negative for the N\'eel (AFM) phase and becomes positive in the Kitaev spin liquid (KSL) phase.
 (c) For $J = 1-\alpha$ and $K =-2\alpha$, the binding energy is negative only in the N\'eel (AFM) phase and becomes positive for the stripy and Kitaev spin liquid (KSL) phases.
	}
\label{fig:binding:alpha}
\end{figure}


We begin by calculating the two-hole binding energy in order to understand the pairing tendencies of holes in the honeycomb Kitaev-Heisenberg model. The binding energy near half-filling is defined as 
\begin{equation}
\Delta E= E(N-2)+E(N)-2E(N-1), \label{eq:binding}
\end{equation}
where $E(n_e)$ is the ground state energy with $n_e$ electrons present. Alternatively, the filling can be quantified by the number of holes $n_h$ introduced on top of the half-filled background, i.e., $n_h=N-n_e$.
Negative values ($\Delta E<0$) indicate that it is energetically favorable for holes to form bound pairs. Such a pairing tendency is a necessary but not sufficient condition for  superconductivity to occur. In contrast, $\Delta E\gtrsim 0$ indicates the absence of bound hole pairs. (Note that $\Delta E$ can be positive in finite-size systems.)

We first investigate the influence of the hopping $t$ on the binding energy in Fig. \ref{fig:binding:hopping}(a). By keeping the system size $N=2\times 3 \times 4$ fixed and varying $t$ we see a strikingly different behavior at the AFM Heisenberg point compared to the two Kitaev points. At the Heisenberg point, the binding energy remains finite and negative over the entire studied range $0\leq t/J\leq 3$. This behavior is consistent with results for the doped $t-J$ model on the square lattice~\cite{PhysRevB.49.12318, RevModPhys.66.763, PhysRevB.94.155120}. In contrast, at the Kitaev points the binding energy is found to be negative only for slow holes with $t<1/4$. At $t=0$, $\Delta E$ is independent of the sign of $K$. 
In this limit, the DMRG calculation is significantly affected by edge effects. Nevertheless, the qualitative result is intuitively clear: formation of hole pairs is expected to be energetically favored as $t\rightarrow 0$ in order to minimize the number of broken magnetic bonds. 
We later show via exact calculations of the Kitaev model that the binding energy at $t=0$ is indeed negative in an infinite system, but smaller in magnitude than predicted by DMRG; see the Discussion in Sec.~\ref{sec:discussion} for details.
For $t>1/4$ the binding energy is positive and appears to grow linearly with $t$, with higher slope for the FM Kitaev point than for the AFM Kitaev point. The difference in slope is consistent with the different tendencies to Nagaoka ferromagnetism noted in the literature \cite{Kadow_2024, Jin2024}. The crossover at $t_c=1/4$ supports the distinction that has been made between fast and slow holes in the Kitaev model \cite{PhysRevB.94.235105, PhysRevB.90.024404}. 

Next, to understand the system-size dependence of this crossover we calculate the binding energy at $K=2,\,J=0$ for different system sizes ($N=24,\,36$ and $48$) and identify the crossover hopping $t_c$, defined as the value of $t$ where $\Delta E$ changes sign. The resulting values are plotted in Fig.~\ref{fig:binding:hopping}(b). From a finite-size extrapolation (using a linear fit of the data), we obtain $t_c\sim 0.8$ in the limit $N\rightarrow \infty$. We note that, due to the limited cylinder length and circumference, a different value could be found in the 2D thermodynamic limit ($L_x\rightarrow \infty$, $L_y\rightarrow \infty$). However, using this value for the finite-size-scaled crossover hopping, our results suggest that, in the spin-liquid regime, pair formation will only occur at relatively small hopping $t_c/|K|\lesssim 0.4$, namely for relatively slow holes. This is a result not sufficiently emphasized in the previous Kitaev model literature.

We proceed by fixing $t=1$, which optimizes the binding energy at the AFM Heisenberg point, and study the behavior of $\Delta E$ in the three different parameter regions shown in Fig. \ref{fig:model} and described in the previous section. We use here three different cluster sizes $N=2\times 3 \times 4 = 24$,  $N=2\times 3 \times 8 = 48$ and  $N=2\times 3 \times 12 = 72$. Figure \ref{fig:binding:alpha}(a) shows the binding energy for $J=\alpha-1\leq 0$ and $K=2\alpha\geq 0$ in the range of $0\leq \alpha \leq 1$. We find that the binding energy remains positive in the FM phase for all system sizes studied. In the zigzag phase, $\Delta E$ is close to zero for the largest systems. In the spin liquid phase (KSL$^1$), where convergence is difficult to achieve, we find that the binding energy remains finite and positive at the largest size, consistent with the behavior in Fig. \ref{fig:binding:hopping}.

Figure \ref{fig:binding:alpha}(b) shows the binding energy for $J=1-\alpha\geq 0$ and $K=2\alpha\geq 0$ with $0\leq \alpha \leq 1$. In this range, only two phases are present: the Kitaev spin liquid and the N\'eel AFM phase. We find negative binding energy only in the latter, with binding optimized at the Heisenberg point and $|\Delta E|$ slowly decreasing as the Kitaev point is approached. Whereas the system-size dependence is notable in the spin liquid phase, the results in the AFM phase for clusters of sizes $N=48$ and $N=72$ are almost the same, indicating a more rapid convergence in cylinder length.

Figure \ref{fig:binding:alpha}(c) shows the behavior for $J=1-\alpha\geq 0$ and $K=-2\alpha\leq 0$ with $0\leq \alpha \leq 1$, including the AFM phase, the stripy phase, and the FM Kitaev point with the KSL$^2$ spin liquid phase. As expected, the binding energy is negative in the N\'eel phase for $\alpha \leq 0.4$. The binding energy approaches zero in the stripy phase, signaling the absence of two-hole bound pairs. (We note an anomaly at $\alpha=0.6$ where the binding energy for $N=72$ is larger than that for $N=48$. This is caused by a numerical convergence issue, but does not affect the physical conclusion.) The binding energy in the spin liquid phase remains positive for the largest system size, and at fixed system size it is larger than at the AFM Kitaev point. Together, these results imply that at this fixed value of hopping ($t=1$), the N\'eel phase stands out as having robust pairing tendencies, while no evidence of pairing is found in the other phases.

\begin{figure}[ht]
\begin{overpic}[width=\columnwidth]{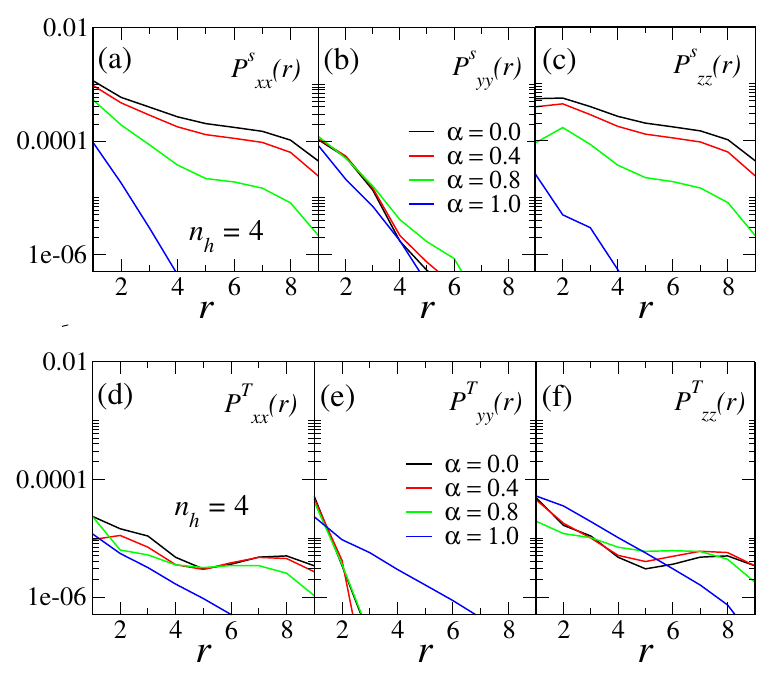} 
\end{overpic} 
\caption{Averaged pair-pair correlations in the $J\geq 0$, $K\geq 0$ sector for different values of $\alpha$ ($J=1-\alpha$ and $K=2\alpha$) at hole doping $n_h=4$ and clusters with $N=72$ sites. The top row shows singlet pair-pair correlation functions along different bond directions: (a) $P^S_{xx}(r)$ along $x$-bonds with distance $r$, (b) $P^S_{yy}(r)$ along $y$-bonds, and (c) $P^S_{zz}(r)$ along $z$-bonds.
The bottom row shows triplet pair-pair correlations: (d) $P^T_{xx}(r)$ along $x$-bonds, (e) $P^T_{yy}(r)$ along $y$-bonds, and (f) $P^T_{zz}(r)$ along $z$-bonds.
}
\label{fig:paircorr:alpha}
\end{figure}

\subsection{Pair-pair correlations}\label{sec:results:pairpair}
\begin{figure}[ht]
\centering
\includegraphics*[width=\columnwidth]{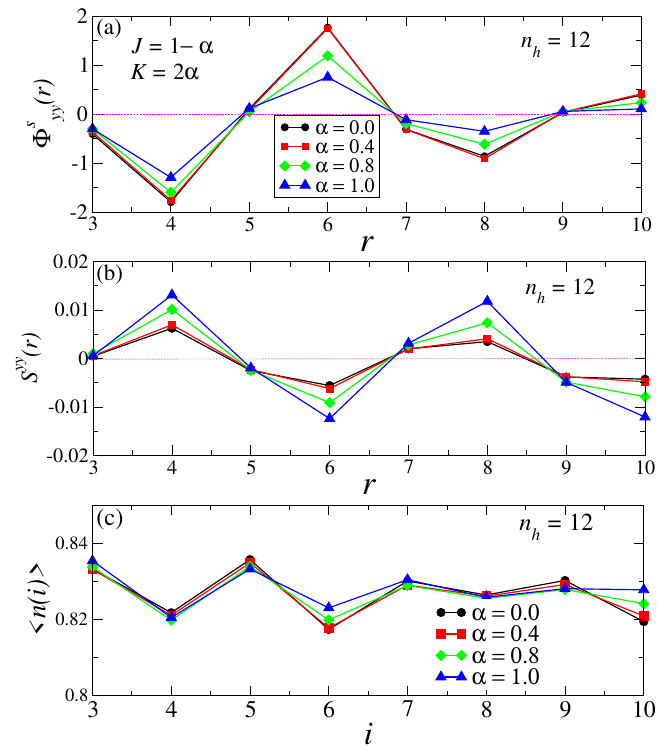}
\caption{(a) Normalized singlet pair-pair correlation 
  $\Phi^S_{yy}(r)=p^S_{yy}(r)/f(r)$ along $y$-bonds of the honeycomb lattice with distance $r$.
(b) $y$-component of the spin-spin correlation $S_{yy}$ along the $\mathbf{a}_1$-direction of the honeycomb lattice.
(c) Rung charge density $\langle n(i)\rangle=\sum_j \langle n(i,j) \rangle /L_y$.
These results were calculated using DMRG at $n_h = 12$ and for different values
of $\alpha$ [with $J = 1-\alpha$ and $K = 2\alpha$]. We have used system size $N= 2\times 3\times 12$
and number of electrons $n_e=60$.
}
\label{fig:pdw}
\end{figure}
Having established that the binding energy calculations predict hole pairing only in the N\'eel phase for the intermediate hopping $t=1$, we next focus on signatures of superconductivity and the nature of pairing in the $J\geq 0$, $K\geq 0$ region ($J=1-\alpha$, $K=2\alpha$). To investigate these aspects, we calculate singlet and triplet pair-pair correlations using $72$-site three-leg cylinders. The pair-pair correlation function can be defined as
\begin{equation}
    p^{S/T}_{\mu \nu}(r)=\langle \Delta_{\mu}^{\dagger}(i_0,j_0) \Delta_{\nu}^{\phantom{\dagger}}(i_0+r,j_0)\rangle,
\end{equation}
where $r$ is the distance along the long direction of the cylinder, $i_0$ and $j_0$ are reference positions along the respective lattice directions labeled $\mathbf{a}_1$ and $\mathbf{a}_2$ in Figure \ref{fig:model}, and
\begin{equation}
    \Delta^{\dagger}_{\mu}(i,j)=\frac{1}{\sqrt{2}} \left [ c^{\dagger}_{(i,j),\uparrow} c^{\dagger}_{(i,j)+\vec{\mu},\downarrow} \pm c^{\dagger}_{(i,j),\downarrow} c^{\dagger}_{(i,j)+\vec{\mu},\uparrow}\right ]
\end{equation} 
is the pair creation operator, with a $-$ sign for singlet pairs and $+$ sign for triplet pairs. The index $\mu \in \{ x, y,z\}$ denotes the type of bond, as illustrated in Fig. \ref{fig:model}, which is associated with a displacement given by the vector $\vec{\mu}$. This 
amounts to an assumption that pairs are predominantly formed from holes on neighboring sites, as is typically the case in $t-J$-like models with short-range interactions, where nearest-neighbor pairs minimize the energy cost incurred from the magnetic part of the Hamiltonian.

Figure \ref{fig:paircorr:alpha} compares the averaged pair-pair correlation functions along different bond directions for four holes as $\alpha$ is varied. The averaging is defined as
\begin{equation}
	 P^{S/T}_{\mu \nu}(r)=\frac{1}{N_r} \sum_{i_0} p^{S/T}_{\mu \nu}(r)
\end{equation}
where $N_r$ is the number of pairs at distance $r$ along the long direction of the cylinder from the reference site $(i_0,j_0)$. We neglect one bond on each end of the cluster to reduce boundary effects. Note that the zigzag termination of the cluster reduces the maximal distance between $y$ bonds by one from the maximal distance between $x$ (or $z$) bonds.

As shown in Figure \ref{fig:paircorr:alpha}, the singlet (triplet) pair correlations 
$P^S_{xx}$ ($P^T_{xx}$) and  $P^S_{zz}$ ($P^T_{zz}$) have very similar power-law decays with distance, across different values of $\alpha \leq 0.8$. (The correlations at the largest distances, $r=8,9$, show some deviations from power-law behavior due to boundary effects and insufficient averaging.) This behavior is expected due to the spatial isotropy of the Hamiltonian and the similar symmetry of the $x$ and $z$ bonds in the finite zigzag-terminated clusters we consider. On the other hand, the pair-pair correlation along $y$ bonds behaves differently, and decays exponentially at all values of $\alpha$; see Fig. \ref{fig:paircorr:alpha}(b),(e). Throughout the N\'eel phase ($\alpha \leq 0.8$) the singlet pair-pair correlations $P^S_{xx}$ and $P^S_{zz}$ dominate and have the largest values at large distances; see Fig. \ref{fig:paircorr:alpha}(a),(c). At the AFM Kitaev point ($\alpha=1$, or $J=0$ and $K=2$), we find that all considered pair-pair correlation functions decay exponentially. This is consistent with our binding energy results.

\subsection{Pair-density-wave}\label{sec:results:PDW}

    \begin{figure}[t]
     \begin{overpic}[width=\columnwidth]{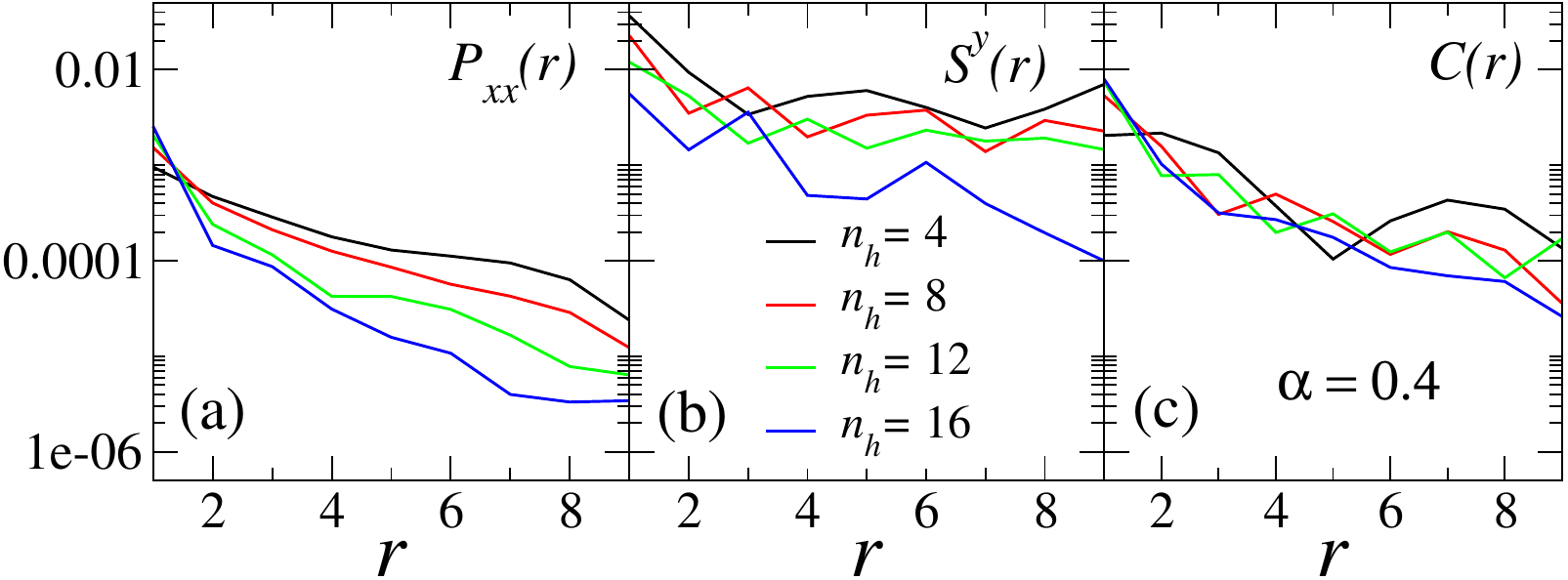}
    \end{overpic}
\caption{Comparison of correlation functions at $\alpha=0.4$ $[J=0.6, K=0.8]$ on clusters with $N=72$ sites for different numbers of introduced hole dopants $n_h$. (a) Singlet pair-pair correlations. (b) Spin-spin correlations. (c) Charge-charge correlations.
}
\label{fig:correlations:alpha0p4}
\end{figure}

In light of the observation of precursors to pair-density wave (PDW) formation at the Kitaev point \cite{Peng2021}, we next explore the possibility of PDW formation in the $J\geq 0$, $K\geq 0$ region of the Kitaev-Heisenberg model. The PDW state is characterized by intertwined superconducting, spin, and charge orders~\cite{RevModPhys.87.457,PhysRevB.79.064515}. In this state, the superconducting (SC) order and spin order have the same periodic modulation, 
while the charge order varies with half the period of the SC and spin orders~\cite{PhysRevLett.105.146403}. The SC order in PDW states also oscillates periodically such that its spatial average vanishes~\cite{PhysRevB.107.214504,PhysRevB.79.064515}.
Figure \ref{fig:pdw}(a) shows the normalized singlet pair-pair correlation 
\begin{equation}
    \Phi^S_{yy}(r)=p^S_{yy}(r)/f(r)
\end{equation}
with pairs along $y$-bonds for $n_h=12$ holes. To highlight the clear spatial oscillations, we fit $f(r)$ to a power-law $f(r)\sim r^{-K_\mathrm{sc}}$ as described in Ref.~\cite{Peng2021}. Note that here $\Phi^S_{yy}(r)$ is \emph{not} averaged over different bonds. Instead, we calculate the pair-pair correlation function away from a fixed site, after neglecting boundary sites as before. 

Interestingly, the normalized pair-pair correlation [Fig. \ref{fig:pdw}(a)] shows spatial oscillations with distance $r$; $\Phi^s_{yy}(r)$ changes sign after each distance of $\approx 2$, such that the spatial average almost vanishes. The spin-spin correlation
\begin{equation}
    S^{yy}(r)   = \langle S^y_{i_0,j_0} S^y_{i_0+r,j_0} \rangle 
\end{equation}
is plotted in Fig. \ref{fig:pdw}(b). Here we chose the reference site at $(i_0,j_0)$ after neglecting
one bond along the left end of the cluster. The spin-spin correlation also oscillates with approximately the same period $2$, but the oscillation is in antiphase with that of the normalized pair-pair correlations. On the other hand, the rung charge density
\begin{equation}
    \langle n(i)\rangle=\sum_j \langle n(i,j) \rangle /L_y
\end{equation}
shows a spatial modulation with a period that is one half of that for the pair-pair and spin-spin correlations ($\approx 1$ site). These results are characteristic of a striped pair-density wave \cite{Peng2021,RevModPhys.87.457,PhysRevB.79.064515}. We note that this type of intertwined spin, charge, and superconducting order is quite distinct from normal coexisting superconducting and charge-density wave orders, where the averaged pair-pair correlations do not vanish and the modulation periods of the pair-pair correlations and the charge density are the same.

As may be expected from the binding energy trends, the pair-density wave signatures become clearer and more dominant while moving from the AFM Kitaev limit ($\alpha=1$) to the AFM Heisenberg limit ($\alpha=0$). What is the symmetry of the pairing process? To investigate this, we calculate the ratio $P_{xy}(r)/P_{yy}(r)$ for different values of $\alpha$. For $\alpha \geq 0.8$, i.e., close to the AFM Kitaev point, we find that the ratio is always negative, indicating a dominant $d$-wave symmetry of the singlet pair-pair correlations. This is consistent with the previously published result at $\alpha=1$ \cite{Peng2021}. On the other hand, for $\alpha \leq 0.6$ we find some oscillations in the sign that could suggest a different pairing symmetry. However, due to the restriction to finite-size clusters we are not able to conclusively determine the nature of the pairing. We note that prior work on the honeycomb $t-J$ model (i.e., $\alpha=0$) reported $d$-wave pairing symmetry, possibly of the time-reversal-breaking $d+id$ variety \cite{PhysRevB.88.155112, PhysRevB.108.035144, Cui2024tJ}.

\subsection{Comparison of correlation functions}\label{sec:results:comparison}

Finally, to determine the dominant instability in the presence of both Kitaev $K$ and Heisenberg $J$ terms, we compare the pair-pair, spin-spin, and charge-charge correlations as functions of the distance $r$ and for different doping levels at $\alpha=0.4$ [$J=0.6$ and $K=0.8$]. We use averaged correlation functions and compare results for $n_h=4,8,12,16$ holes, corresponding to hole doping concentrations of $n_h/N=5.6\%,11\%,17\%,22\%$. The charge-charge correlation function is defined as:
\begin{align}
    C(r)    &= \frac{1}{N_r} \sum_{i_0}\left\{ \langle n(i_0,j_0) n(i_0+r,j_0) \rangle\right. \nonumber\\
            &\left. - \langle n(i_0,j_0) \rangle  \langle n(i_0+r,j_0) \rangle \right\}
\end{align}
 The results are plotted in Fig. \ref{fig:correlations:alpha0p4}. We find that the averaged spin-spin correlation $S^y(r)=\frac{1}{N_r}  \sum_{i_0}  S^{yy}(r)$ dominates for all hole concentrations. This is expected due to the odd number of ``legs'' in the finite-size cluster, and is reflective of the even-odd effect also seen in square-lattice ladders \cite{doi:10.1126/science.271.5249.618, Zhu2014_evenodd}. We expect $P_{xx}(r)$, $S^y(r)$, and $C(r)$ to compete more closely in cylinders of width $L_y=4$, and more generally for $L_y$ even. As the hole doping is increased, we find that the decay rate of all correlation functions increase. This is particularly notable in the spin-spin correlations as $n_h$ is increased from $4$ to $16$, presumably because of increasingly severe scrambling of the spin order by the additional introduced holes~\cite{PhysRevB.103.214513}.

\section{Discussion}\label{sec:discussion}

Our results on the binding energy show that prior phase diagrams obtained from slave-boson mean-field theory \cite{PhysRevB.85.140510, PhysRevB.86.085145, PhysRevB.87.064508, PhysRevB.97.014504} neglect crucial effects relating to the kinetic energy of the doped holes. In the quasi-stationary regime ($t\ll 1$), the holes can be viewed as spin vacancies, and the formation of hole pairs is readily understood from a simple intuitive argument following Refs.~\onlinecite{PhysRevB.49.12318, RevModPhys.66.763}. Since each hole breaks three bonds of the honeycomb lattice, it has a finite energy cost due to the three associated magnetic interactions. For two holes, it is then energetically favorable to be on neighboring sites so that they only break five bonds instead of six. While this argument was originally presented for the N\'eel phase on the square lattice~\cite{PhysRevB.49.12318, RevModPhys.66.763}, it straightforwardly generalizes to all phases in this work, including the Kitaev spin liquids.

\begin{figure}[hbt]
\centering
\includegraphics*[width=\columnwidth]{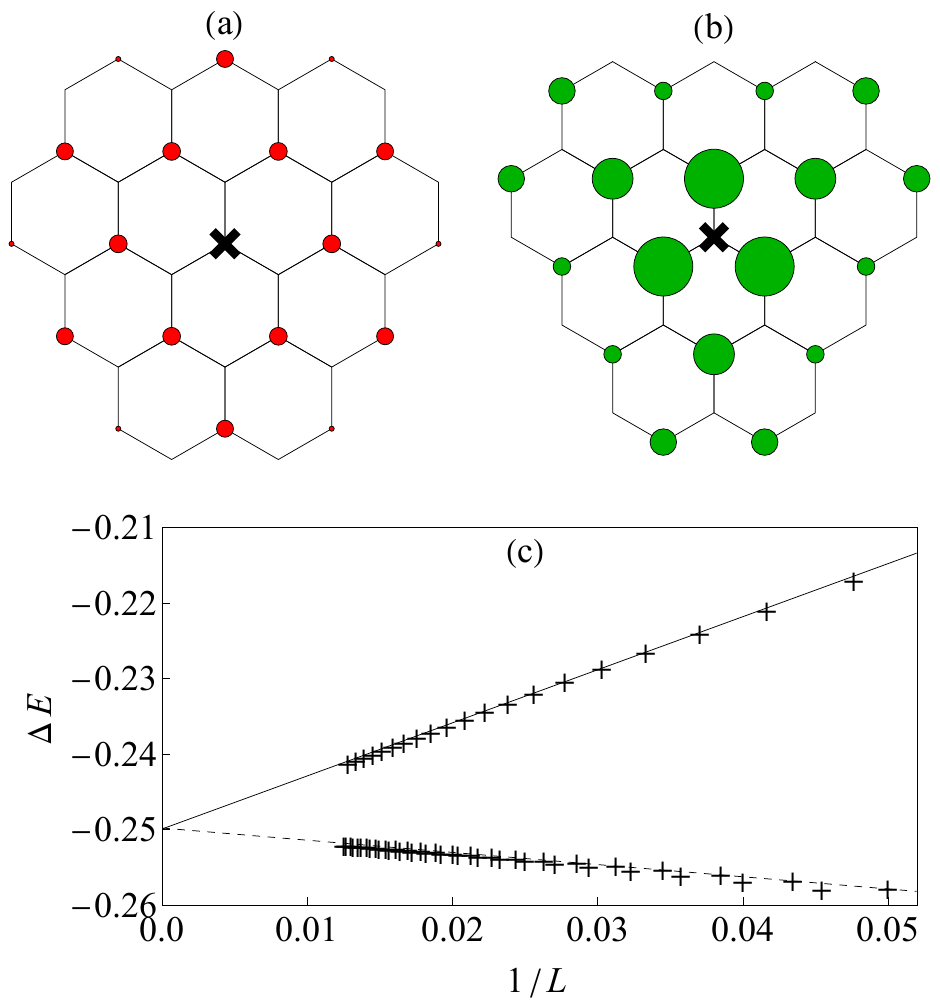}
\caption{(a,b) Binding energy of two holes in the $t\to 0$ limit for the pure Kitaev model with $|K|=2$ as a function of their relative position if the holes are on the same sublattice (a) or opposite sublattices (b). One hole is fixed (black cross), and the other one is moved around (colorful disks). The area of each disk is proportional to the absolute value of the binding energy, $|\Delta E|$, while the color of the disk is red for $\Delta E>0$ and green for $\Delta E<0$. The binding energies are calculated for an $L \times L$ honeycomb lattice with $L=40$. (c) Finite-size scaling of the nearest-neighbor binding energy. The solid and dashed lines are linear extrapolations from data points with $L\,\mathrm{mod}\,3 = 0$ and $L\,\mathrm{mod}\,3 \neq 0$, respectively.}
\label{fig:exact}
\end{figure} 

Indeed, one can explicitly confirm pair formation in the pure Kitaev model, which remains exactly solvable even in the presence of spin vacancies~\cite{PhysRevLett.104.237203, PhysRevB.84.115146, PhysRevB.90.035145}. For the Kitaev model on a $40 \times 40$ honeycomb lattice, the binding energy of two nearby vacancies, as defined by Eq.~(\ref{eq:binding}), is plotted against relative position in Figs.~\ref{fig:exact}(a,b). In general, the binding energy is positive (negative), corresponding to repulsion (attraction), if the two vacancies are on the same sublattice (opposite sublattices). Since the most negative binding energy is found for the nearest-neighbor position, it is indeed energetically favorable for two vacancies to be on neighboring sites. From the finite-size scaling shown in Fig.~\ref{fig:exact}(c), we then determine the nearest-neighbor binding energy to be $-0.250$ for $|K|=2$. If we also account for the gauge fluxes attached to isolated vacancies in the Kitaev model, lowering the energy of each well-separated pair by $-0.027$~\cite{PhysRevLett.104.237203}, the binding energy of two vacancies is finally found to be $-0.223$. This binding energy is around one half of the one found in Fig.~\ref{fig:binding:hopping}(a); the most probable sources of the discrepancy are the finite cylinder width and the open ends of the cylinder in DMRG. Nevertheless, the binding of vacancies is qualitatively well captured by DMRG. We remark that the binding energy of two vacancies is the same for the FM and AFM Kitaev models. Moreover, the position dependence in Figs.~\ref{fig:exact}(a,b) is similar to the anisotropic gapped phase of the Kitaev model~\cite{PhysRevB.90.035145}, although with a slower decay, as expected for a gapless phase.

In realistic hole-doped bulk quantum materials, we expect faster holes corresponding to $t\geq 1$. Thus, since our results only show binding at the Kitaev points for slow holes, and no binding in the FM, zigzag, stripy, and KSL phases for $t=1$, pairing in real bulk materials described by the Kitaev-Heisenberg model can likely only be found in the N\'eel phase. However, firm conclusions about the behavior in the thermodynamic limit cannot be drawn until finite-size scaling in the cylinder width (which is computationally very challenging) can be achieved. We also note that
the description of existing Kitaev candidate materials like $\alpha$-RuCl$_3$ and Na$_2$IrO$_3$ requires additional interactions, such as off-diagonal $\Gamma$ interactions and further-range Heisenberg exchange, which are known to dramatically modify the phase diagram at half-filling \cite{PhysRevLett.112.077204}. What effects they might have on the pairing tendencies is presently unclear, leaving open the possibility that they might stabilize pairing over a broader parameter range. Thus, future extensions of the current study to, for example, the Kitaev-Heisenberg-$\Gamma$ model are encouraged.

A more promising path to accessing the slow-hole regime could be heterostructures involving exfoliated layers of Kitaev systems in proximity to other materials. For example, $\alpha$-RuCl$_3$/graphene heterostructures have been reported to induce light electron doping in the Kitaev layer and to enhance the Kitaev interaction strength, leading to an estimated $t/|K|\approx 0.4$ \cite{PhysRevLett.123.237201}. This ratio is very similar to our estimate for the crossover hopping in the Kitaev limit, $t_c/|K|\sim 0.4$. However, it must be noted that the spin Hamiltonian used in Ref. \cite{PhysRevLett.123.237201} includes additional spin-spin interactions, meaning a direct comparison cannot be made. We encourage further study of such heterostructures. 
In addition, we note that the slow-hole regime may be accessible to quantum simulation platforms \cite{Bohrdt2021, PRXQuantum.4.020329, PhysRevLett.132.186501}.

Finally, the appearance of pair-density wave behavior over a broad range of parameters in the Kitaev-Heisenberg model is interesting and promising for further study. In particular, it means that this behavior remains, and is in fact strengthened, away from the numerically challenging Kitaev point, where pair-density wave precursors have already been found \cite{Peng2021}. It also locates our system in the company of models with PDWs on triangular lattices \cite{PhysRevLett.122.167001, Peng_2021} and a spinless fermion model on the honeycomb lattice \cite{PhysRevLett.133.176501}.

\section{Conclusion}\label{sec:conclusion}
We have studied the pairing tendencies in a hole-doped Kitaev-Heisenberg model, the $t-J-K$ model, numerically using DMRG. In the Kitaev limit ($J=0$) we find that the binding energy between two holes depends strongly on the hopping $t$ that determines the kinetic energy of the holes. This is in agreement with the distinction between slow and fast holes that has previously been made in analytical and computational studies on the one-hole problem. As a result, we only predict pair formation at relatively low hopping, $t/|K| \lesssim 0.4$, a regime of potential relevance to heterostructures but unlikely to occur in bulk Kitaev candidate materials.

In contrast, in the Heisenberg limit ($K=0$) the binding energy indicates pair formation up to at least $t/J\geq 3$, in qualitative agreement with the behavior of the square lattice $t-J$ model \cite{RevModPhys.66.763}. Hence, for fixed $t=1$, $|J|\leq 1$ and $|K|\leq 2$ we only find evidence of pair formation in the N\'eel phase. These results, while subject to finite-size limitations, suggest that the dynamical effects of holes play an important role in doped Kitaev-Heisenberg models, and that current slave-boson mean field theories may need to be extended. We have also found that the pair-density wave behavior previously reported in the Kitaev limit in fact extends over a range of parameters in the Kitaev-Heisenberg model.

\begin{acknowledgments}
The work of PL, BP, SO, and ED was supported by the U.S. Department of Energy (DOE), Office of Science, Basic Energy Sciences (BES), Materials Sciences and Engineering Division. The work of GBH and GA was supported by the U.S. Department of Energy, Office of Science, National Quantum Information Science Research Centers, Quantum Science Center.
\end{acknowledgments}

%


\newpage
\clearpage
\title{Supplemental material for ``Pairing tendencies in the doped Kitaev-Heisenberg model''}
\maketitle
\onecolumngrid

\section*{Reproducing the DMRG results.}
Here we describe how the DMRG results displayed in the main text can be reproduced. This work used DMRG++ versions 6.05--6.07 and PsimagLite versions 3.04--3.07.

\subsection{Building DMRG++}
The \textsc{DMRG++} computer program may be obtained with:
\begin{verbatim}
git clone https://github.com/g1257/dmrgpp.git
\end{verbatim}
The BOOST and HD5 libraries are required dependencies. As is PsimagLite, which can be obtained with:
\begin{verbatim}
git clone https://github.com/g1257/PsimagLite.git
\end{verbatim}
To compile:
\begin{verbatim}
cd PsimagLite/lib; perl configure.pl; make
cd ../../dmrgpp/src; perl configure.pl; make
\end{verbatim}
To simplify commands below we also run
\begin{verbatim}
export PATH="<PATH-TO-DMRG++>/src:$PATH"
export SCRIPTS="<PATH-TO-DMRG++>/scripts"
\end{verbatim}

The documentation can be found at 
\nolinkurl{https://g1257.github.io/dmrgPlusPlus/manual.html} Alternatively, it can be built locally by doing
\begin{verbatim}cd ../..; git clone https://github.com/g1257/thesis.git; cd dmrgpp/doc; ln -s ../../thesis/thesis.bib; 
make manual.pdf
\end{verbatim}

\subsection{Templated ground state inputs}
DMRG++ provides support for cylinder geometries through inputs that express the Hamiltonian as a set of matrices corresponding to two different two-site terms (called connections). These can be generated directly by the user, or, more conveniently, by scripts. For this purpose, DMRG++ includes \verb!honeycomb.pl!, which converts templated inputs into inputs with explicit matrices. The script can handle both symbolic and numerical values for coupling constants. An example symbolic templated input \verb*|InputTemplateSymbolic.din| is given by
\begin{verbatim}
TotalNumberOfSites=$n
NumberOfTerms=$nOfTerms
GeometryMaxConnections=3

Model=KitaevWithCharge
SolverOptions=twositedmrg
Version=symbolic
OutputFile=dataGS
InfiniteLoopKeptStates=200
FiniteLoops 5
$halfNminusOne 300 0
-$nMinusTwo 400 0     $nMinusTwo 400 0
-$nMinusTwo 800 0     $nMinusTwo 800 1

TruncationTolerance=1e-7
TargetElectronsTotal=$n

$replaceConnections

DefineOperators=nup:c'*c,ndown:c?1'*c?1,n:nup+ndown
#lx=2
#ly=2
#options=periodicy,zigzag
#Hoppings=-tx,-ty,-tz
#kxkx=Kx
#kyky=Ky
#kzkz=Kz
#jx=J
#jy=J
#jz=J
\end{verbatim}
where the number of sites is determined from the multiples of the lattice vectors, i.e. \$n=2$\times$lx$\times$ly. The option \verb!periodicy! indicates periodic boundary conditions along the second lattice vector, and \verb!zigzag! indicates a zigzag termination. For cylinder geometries, the value of \verb!GeometryMaxConnections! should be $\geq2$ly$-1$, with equality being optimal to minimize memory usage. Alternatively, the value could be set to $0$ to keep in memory connections between all sites. Each finite loop is associated with a triplet of values, where the last value being set to $1$ indicates that DMRG++ should write to disk all information required to measure correlation functions.

Doing \verb!perl -I ${SCRIPTS} ${SCRIPTS}/honeycomb.pl InputTemplateSymbolic.din! generates a .inp input file of the form
\begin{verbatim}
TotalNumberOfSites=8
NumberOfTerms=4
GeometryMaxConnections=3

Model=KitaevWithCharge
SolverOptions=twositedmrg
Version=dopedKitaev
OutputFile=dataGS
InfiniteLoopKeptStates=200
FiniteLoops 5
3 300 0
-6 400 0     6 400 0
-6 800 0     6 800 1

TruncationTolerance=1e-7
TargetElectronsTotal=8

GeometryKind=LongRange
GeometryOptions=none
GeometryMaxConnections=0
Connectors 8 8
0  -tx  0  -tz  0  0  0  0 
0  0  -tz  0  -ty  0  0  0 
0  0  0  -tx  0  0  0  0 
0  0  0  0  0  0  -ty  0 
0  0  0  0  0  -tx  0  -tz 
0  0  0  0  0  0  -tz  0 
0  0  0  0  0  0  0  -tx 
0  0  0  0  0  0  0  0 



GeometryKind=LongRange
GeometryOptions=none
GeometryMaxConnections=0
Connectors 8 8
0  Kx + J  0  J  0  0  0  0 
0  0  J  0  J  0  0  0 
0  0  0  Kx + J  0  0  0  0 
0  0  0  0  0  0  J  0 
0  0  0  0  0  Kx + J  0  J 
0  0  0  0  0  0  J  0 
0  0  0  0  0  0  0  Kx + J 
0  0  0  0  0  0  0  0 


GeometryKind=LongRange
GeometryOptions=none
GeometryMaxConnections=0
Connectors 8 8
0  J  0  J  0  0  0  0 
0  0  J  0  Ky + J  0  0  0 
0  0  0  J  0  0  0  0 
0  0  0  0  0  0  Ky + J  0 
0  0  0  0  0  J  0  J 
0  0  0  0  0  0  J  0 
0  0  0  0  0  0  0  J 
0  0  0  0  0  0  0  0 


GeometryKind=LongRange
GeometryOptions=none
GeometryMaxConnections=0
Connectors 8 8
0  J  0  Kz + J  0  0  0  0 
0  0  Kz + J  0  J  0  0  0 
0  0  0  J  0  0  0  0 
0  0  0  0  0  0  J  0 
0  0  0  0  0  J  0  Kz + J 
0  0  0  0  0  0  Kz + J  0 
0  0  0  0  0  0  0  J 
0  0  0  0  0  0  0  0 



DefineOperators=nup:c'*c,ndown:c?1'*c?1,n:nup+ndown
#lx=2
#ly=2
#options=periodicy,zigzag
#Hoppings=-tx,-ty,-tz
#kxkx=Kx
#kyky=Ky
#kzkz=Kz
#jx=J
#jy=J
#jz=J

\end{verbatim}
showing the structure of the connection matrices. Note that only the upper triangular parts are nonzero. In order, the matrices represent hopping, $S^xS^x$, $S^yS^y$, and $S^zS^z$ interactions.

\begin{figure}
	\includegraphics{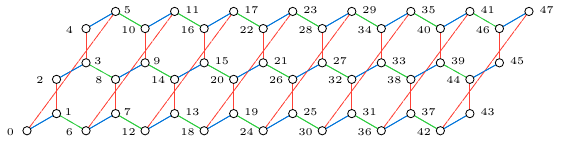}
	\label{fig:sm:lattice}
	\caption{48-site cluster with $L_x=8$, $L_y=3$. Dashed red lines indicate sites connected by the periodic boundary conditions.}
\end{figure}

We also provide an example templated input with numerical values and 48 sites, \verb!InputTemplate.din!:
\begin{verbatim}
TotalNumberOfSites=$n
NumberOfTerms=$nOfTerms
GeometryMaxConnections=5

Model=KitaevWithCharge
SolverOptions=twositedmrg
Version=dopedKitaev
OutputFile=dataGS
InfiniteLoopKeptStates=200
FiniteLoops 33
$halfNminusOne 300 0
-$nMinusTwo 400 0     $nMinusTwo 400 0
-$nMinusTwo 800 0     $nMinusTwo 800 0
-$nMinusTwo 1200 0     $nMinusTwo 1200 0
-$nMinusTwo 1600 0     $nMinusTwo 1600 0
-$nMinusTwo 2000 0     $nMinusTwo 2000 0
-$nMinusTwo 2400 0     $nMinusTwo 2400 0
-$nMinusTwo 2800 0     $nMinusTwo 2800 0
-$nMinusTwo 3000 0     $nMinusTwo 3000 0
-$nMinusTwo 3400 0     $nMinusTwo 3400 0
-$nMinusTwo 3800 0     $nMinusTwo 3800 0
-$nMinusTwo 4000 0     $nMinusTwo 4000 0
-$nMinusTwo 4200 0     $nMinusTwo 4200 0
-$nMinusTwo 4400 0     $nMinusTwo 4400 0
-$nMinusTwo 4600 0     $nMinusTwo 4600 0
-$nMinusTwo 4800 0     $nMinusTwo 4800 0
-$nMinusTwo 5000 0     $nMinusTwo 5000 1

TruncationTolerance=1e-7
TargetElectronsTotal=20
Threads=16

$replaceConnections

DefineOperators=nup:c'*c,ndown:c?1'*c?1,n:nup+ndown
#lx=8
#ly=3
#options=periodicy,zigzag
#Hoppings=-1,-1,-1
#kxkx=0.8
#kyky=0.8
#kzkz=0.8
#jx=0.6
#jy=0.6
#jz=0.6
\end{verbatim}
where \verb!Threads! should be adapted according to the hardware used. Performing 
\begin{verbatim}
perl -I ${SCRIPTS} ${SCRIPTS}/honeycomb.pl InputTemplateSymbolic.din
\end{verbatim} 
again generates a .inp input file. We can also do 
\begin{verbatim}
perl -I ${SCRIPTS} ${SCRIPTS}/honeycomb.pl InputTemplateSymbolic.din InputTemplate.tex
pdflatex *.tex
\end{verbatim}
which in addition generates a graphical representation of the finite-size cluster, as shown in Fig. \ref{fig:sm:lattice}.

\subsection{Calculations}
To perform the ground state run, do \verb!dmrg -f input.inp!. We can control the behavior further and do, for example,
\begin{verbatim}
dmrg -f input.inp "<gs|n|gs>,<gs|sz|gs>" -p 10
\end{verbatim}
where the \verb!-p! flag fixes the precision of the printed output, and we also perform \emph{in situ} measurements of the charge density and $S^z$ operators. After the ground state run is finished, and has properly output the state to a \verb!.hd5! file, we can use \verb!observe! to measure correlation functions, e.g.
\begin{verbatim}
observe -f input.inp "<gs|sz;sz|gs>" >>sz.dat
observe -f input.inp "<gs|splus;sminus|gs>" >>spsm.dat
observe -f input.inp "<gs|sminus;splus|gs>" >>smsp.dat
observe -f input.inp "<gs|n;n|gs>" >>nnc.dat
observe -f input.inp "<gs|n|gs>" >>nn.dat
\end{verbatim}
for spin-spin and charge density-charge density correlations. For four-point correlation functions, DMRG++ by default calculates values for all sites $i<j<k<l$, which is time consuming. This behavior can be controlled using an ``action'' for variables \%0, \%1, \%2, \%3 corresponding to the four site indices as follows
\begin{verbatim}	
observe -f input.inp "{action=%1==+:%0:1&%3==+:%2:1},<gs|c?0';c?1';c?1;c?0|gs>" >>cucdcdcu.dat
observe -f input.inp "{action=%1==+:%0:1&%3==+:%2:1},<gs|c?0';c?1';c?0;c?1|gs>" >>cucdcucd.dat
observe -f input.inp "{action=%1==+:%0:1&%3==+:%2:1},<gs|c?1';c?0';c?1;c?0|gs>" >>cdcucdcu.dat
observe -f input.inp "{action=%1==+:%0:1&%3==+:%2:1},<gs|c?1';c?0';c?0;c?1|gs>" >>cdcucucd.dat
observe -f input.inp "{action=%1==+:%0:5&%3==+:%2:5},<gs|c?0';c?1';c?1;c?0|gs>" >>cucdcdcu1.dat
observe -f input.inp "{action=%1==+:%0:5&%3==+:%2:5},<gs|c?0';c?1';c?0;c?1|gs>" >>cucdcucd1.dat
observe -f input.inp "{action=%1==+:%0:5&%3==+:%2:5},<gs|c?1';c?0';c?1;c?0|gs>" >>cdcucdcu1.dat
observe -f input.inp "{action=%1==+:%0:5&%3==+:%2:5},<gs|c?1';c?0';c?0;c?1|gs>" >>cdcucucd1.dat
\end{verbatim}
where, in the first four lines we calculate correlation functions with $j=i+1$, $l=k+1$ and in the last four lines we calculate correlation functions with $j=i+5$, $l=k+5$. These choices correspond to the nearest-neighbor pairs in Fig. \ref{fig:sm:lattice}.

\end{document}